\documentclass[twocolumn]{NobArticle}
\usepackage{float}
\usepackage{graphicx}
\usepackage{multirow}

\title{Parallelizing Drug Discovery: HPC Pipelines for Alzheimer’s Molecular Docking and Simulation}
\author{
  Paul Ruiz Alliata\thanks{Corresponding author: pruizalliata.ieu2022@student.ie.edu}$^{1}$,
  Diana Rubaga$^{1}$,
  Daniel Kumlin$^{1}$,
  Alberto Puliga$^{1}$ \\
  $^{1}$Computer Science and Artificial Intelligence Department, IE University, Madrid, Spain
}

\begin{document}
\twocolumn[
  \maketitle

  \begin{abstract}
  High-performance computing (HPC) is reshaping computational drug discovery by enabling large-scale, time-efficient molecular simulations. In this work, we explore HPC-driven pipelines for Alzheimer’s disease drug discovery, focusing on virtual screening, molecular docking, and molecular dynamics simulations. We implemented a parallelised workflow using GROMACS with hybrid MPI–OpenMP strategies, benchmarking scaling performance across energy minimisation, equilibration, and production stages. Additionally, we developed a docking prototype that demonstrates significant runtime gains when moving from sequential execution to process-based parallelism using Python’s multiprocessing library. Case studies on prolinamide derivatives and baicalein highlight the biological relevance of these workflows in targeting amyloid-beta and tau proteins. While limitations remain in data management, computational costs, and scaling efficiency, our results underline the potential of HPC to accelerate neurodegenerative drug discovery.
  \end{abstract}
  \bigskip
]
\section{Introduction}

    \subsection{Alzheimer’s Disease And The Need For New Approaches}
    Alzheimer’s disease (AD) is one of the most prevalent neurodegenerative disorders, 
    accounting for 60–80\% of dementia cases worldwide. It is associated with progressive 
    memory loss and cognitive decline, affecting over 35 million people today, with projections 
    to double in the next 15 years \cite{WHO}. Despite decades of research AD remains extremely difficult to treat, with a 99.6\% clinical 
    trial failure rate \cite{Cummings2014}. This unmet medical need underscores the importance 
    of accelerating drug discovery pipelines with innovative methodologies.  
    
    \subsection{High-Performance Computing In Drug Discovery}
    High-Performance Computing (HPC) enables large-scale simulation and analysis of 
    molecular interactions that are otherwise prohibitively costly or slow in laboratory 
    settings. Techniques such as virtual screening, molecular docking, and molecular 
    dynamics (MD) simulations provide atomistic insights into amyloid-beta (A$\beta$) and tau 
    proteins—key hallmarks of AD pathology. By leveraging parallel processing, GPU 
    acceleration, and optimized algorithms, HPC allows researchers to rapidly screen 
    millions of compounds, evaluate protein–ligand interactions, and model protein 
    aggregation dynamics central to AD progression.  
    
    \subsection{Scope And Contributions Of This Study}
    This paper provides both a review and an implementation-focused study of HPC in 
    Alzheimer’s drug discovery. We benchmarked molecular dynamics simulations of A$\beta$ 
    peptides using GROMACS under hybrid MPI–OpenMP parallelization across three 
    simulation stages: energy minimization, temperature equilibration, and production MD. 
    We further developed a prototype docking algorithm to compare sequential and 
    parallel execution. To ground our computational analysis in biological relevance, we 
    review case studies on prolinamide derivatives and baicalein, compounds investigated 
    for their potential to inhibit AD-related protein aggregation. 
    
    \section{Literature Review }
    
    Alzheimer’s disease (AD) was first described in 1907 by Alois Alzheimer as a “peculiar disease of the cerebral cortex” \cite{Alzheimer1907} and remains one of the most pressing global health challenges. Today, AD accounts for the majority of dementia cases, affecting more than 35 million people worldwide, with projections set to double in the next two decades \cite{WHO}. Despite decades of investment, the AD drug development pipeline has been characterized by exceptionally high failure rates, with an estimated 99.6\% of clinical trials proving unsuccessful \cite{Carroll2014,Cummings2014}. This inefficiency underscores the urgent need for new approaches in therapeutic discovery.  
    
    Computational methods, particularly those enabled by High-Performance Computing (HPC), have become increasingly central to drug discovery. HPC allows researchers to perform large-scale molecular simulations that capture the complex dynamics of amyloid-beta (A$\beta$) and tau proteins implicated in AD pathology \cite{Forman2004,Jack2009,Chakrabarti2015}. Molecular dynamics (MD) simulations, in particular, provide atomistic insights into protein aggregation and destabilization, as demonstrated in early studies \cite{Lemkul2010} and in more recent applications \cite{Ahmed2023,Choudhury2024}. Dedicated HPC architectures such as Anton-2 have further accelerated long-timescale MD, enabling simulations of protein folding and fibril formation at previously unattainable resolutions \cite{Anton2}.  
    
    Parallel advances in virtual screening and molecular docking have also highlighted the role of HPC in accelerating drug discovery. Computational pipelines have identified potential acetylcholinesterase inhibitors \cite{Chennai2024}, JNK3 inhibitors \cite{Devi2024}, and prolinamide derivatives \cite{Olalekan2024} as candidate therapeutics for AD. These studies illustrate how HPC-enhanced docking and MD refinement can uncover multi-targeted compounds capable of addressing the multifactorial nature of the disease. Broader reviews of computational drug discovery approaches for neurodegenerative disorders emphasize the central role of HPC in scaling screening libraries and improving predictive accuracy \cite{Vicidomini2024,Liu2016,Wermuth2015}.  
    
    Beyond conventional HPC, emerging paradigms in big data analytics and artificial intelligence are reshaping AD research. HPC-driven big data frameworks have enabled large-scale integration of genomic and neuroimaging datasets \cite{Chen2019,Jiang2021}, while machine learning models are increasingly applied to dementia prediction and clinical care pathways \cite{Wang2024,WongLin2020}. More recently, exascale computing and quantum technologies have been identified as transformative enablers, capable of handling the unprecedented scale of molecular data and accelerating discovery pipelines \cite{Becker2018,Martin2022,Evers2021,Cappiello2024}.  
    
    Taken together, the literature highlights both the promise and the challenges of computational approaches to AD drug discovery. While HPC has already deepened our understanding of molecular mechanisms and identified promising compounds, the complexity of AD pathology demands continued innovation, including the integration of AI, exascale infrastructures, and quantum-enhanced simulations. This study builds on this foundation by implementing and benchmarking HPC pipelines for molecular dynamics and docking, with specific case studies on prolinamide derivatives and baicalein to illustrate their relevance to Alzheimer’s disease therapeutics.  
    
\section{Methodology}

\subsection{System Setup And Amyloid-\texorpdfstring{$\beta$}{beta} Structure}
The subject of our study was the amyloid-$\beta$ (A$\beta$) peptide, a key biomarker in Alzheimer’s disease. The peptide is strongly implicated in plaque aggregation and neurotoxicity, making it a biologically relevant target for simulation benchmarks. The initial three-dimensional structure of A$\beta$ was obtained and visualized using the \texttt{nglview} library. Atoms were color-coded by type (hydrogen in white, oxygen in red, nitrogen in blue, and carbon in grey) to facilitate structural inspection.

\begin{figure}[H]
  \centering
  \includegraphics[width=0.4\textwidth]{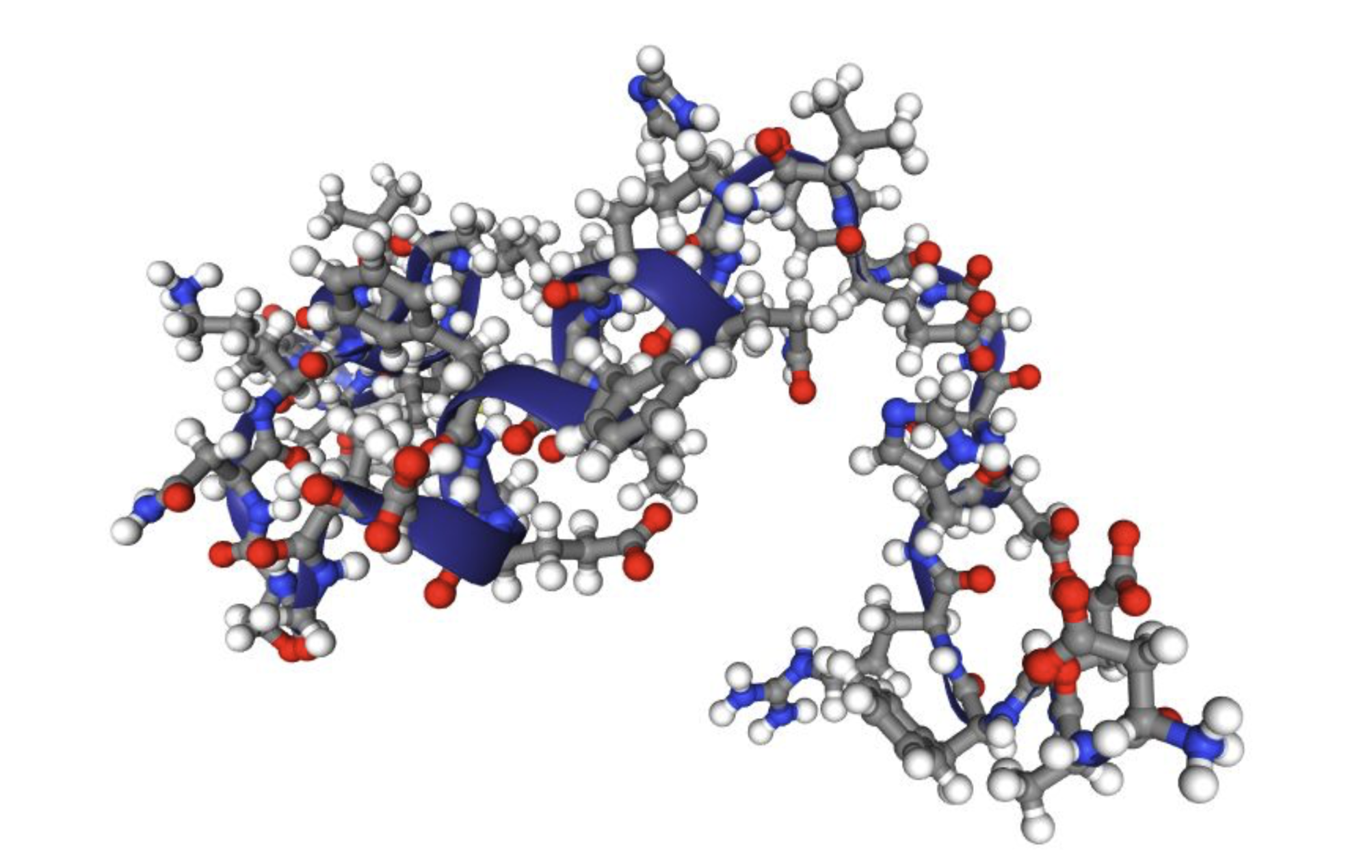}
  \caption{Amyloid-$\beta$ (A$\beta$) peptide structure visualized with \texttt{nglview}. Atoms are color-coded: hydrogen (white), oxygen (red), nitrogen (blue), carbon (grey). This served as the initial configuration for molecular dynamics simulations.}
  \label{fig:abeta}
\end{figure}

\subsection{Computational Pipeline And HPC Optimization}
To evaluate the impact of High-Performance Computing (HPC) on drug discovery workflows, we constructed a molecular dynamics (MD) simulation pipeline optimized for parallel architectures. The simulations were carried out with \texttt{GROMACS} (version 2023.3), an open-source package that integrates highly optimized compute kernels written with SIMD vectorization. \texttt{GROMACS} supports a hybrid parallelization strategy, combining MPI for distributed-memory domain decomposition and OpenMP for shared-memory threading. This model enables scaling across both multi-core CPUs and HPC clusters.

The governing equations follow Newton’s second law of motion:
\[
m \frac{d^2 \mathbf{x}_i}{dt^2} = 
- \sum_{\substack{j=1 \\ j \neq i}}^{N} \frac{\partial \phi_{ij}}{\partial \mathbf{x}_i} 
- \frac{\partial V}{\partial \mathbf{x}_i},
\quad 0 < i \leq N
\]
where:
\begin{itemize}
    \item $m$ is the mass of particle $i$,
    \item $\mathbf{x}_i$ is the 3D position vector of particle $i$,
    \item $\phi_{ij}$ is the pairwise interaction potential between particles $i$ and $j$,
    \item $V$ is the external or collective potential energy.
\end{itemize}

This formulation allows accurate simulation of atomic interactions within biomolecular systems.

\subsection{Simulation Workflow}
The simulation pipeline consisted of three stages: 

\begin{enumerate}
    \item \textbf{Energy Minimization (EM):} Removal of steric clashes and unrealistic geometries from the input structure using the Steepest Descent algorithm. Minimization was terminated either when maximum atomic forces dropped below 1000 kJ/mol/nm or after 50,000 iterations.
    
    \item \textbf{Equilibration (NVT):} Temperature equilibration at 300 K under constant volume (NVT ensemble). The target temperature reflects physiological conditions, ensuring biomolecular relevance and comparability with prior studies.
    
    \item \textbf{Production Simulation (MD):} Long-timescale molecular dynamics runs. We applied the leap-frog integrator for numerical stability, the Parrinello–Rahman barostat for pressure coupling, and the velocity-rescale (V-rescale) thermostat for temperature control. This combination ensured stable trajectories at the target conditions.
\end{enumerate}

\subsection{Parallelization Strategy}
HPC efficiency was tested using a hybrid MPI+OpenMP configuration on a single-node Apple M2 Pro system. We varied the number of OpenMP threads (1–8) while distributing spatial subdomains across MPI ranks. Each MPI process handled a subset of atoms, with inter-process communication ensuring consistent force calculations at domain boundaries. This dual-layer parallelization allowed us to benchmark scaling efficiency, wall-clock time, and parallel overhead across simulation stages (EM, NVT, MD).

\subsection{Docking Pipeline}
To complement MD simulations, a simplified docking prototype was implemented and parallelized. Docking experiments were conducted across increasing numbers of conformations (10, 100, 500). 

\subsection{Benchmarking Metrics}
Performance was evaluated using:
\begin{itemize}
    \item \textbf{Speedup:} Relative runtime improvement with parallel threads.
    \item \textbf{Parallel Efficiency:} Ratio of speedup to the number of threads.
    \item \textbf{Wall-Clock Time:} Real execution time across stages.
\end{itemize}

These benchmarks allowed us to quantify the benefits and limitations of hybrid MPI+OpenMP parallelization in biomolecular workloads.

\section{Results}

We report performance across three MD stages—Energy Minimization (EM), NVT equilibration (NVT), and production MD (MD)—under OpenMP thread counts $p \in \{1,2,4,8\}$. Metrics are: (i) speedup $S(p)=T(1)/T(p)$, (ii) parallel efficiency $E(p)=S(p)/p$, and (iii) wall-clock time $T(p)$. We also evaluate a parallel docking prototype versus a sequential baseline over 10, 100, and 500 conformers.

\subsection{Molecular Dynamics Performance}

\begin{figure}[H]
  \hspace*{-0.5cm} 
  \includegraphics[width=0.50\textwidth]{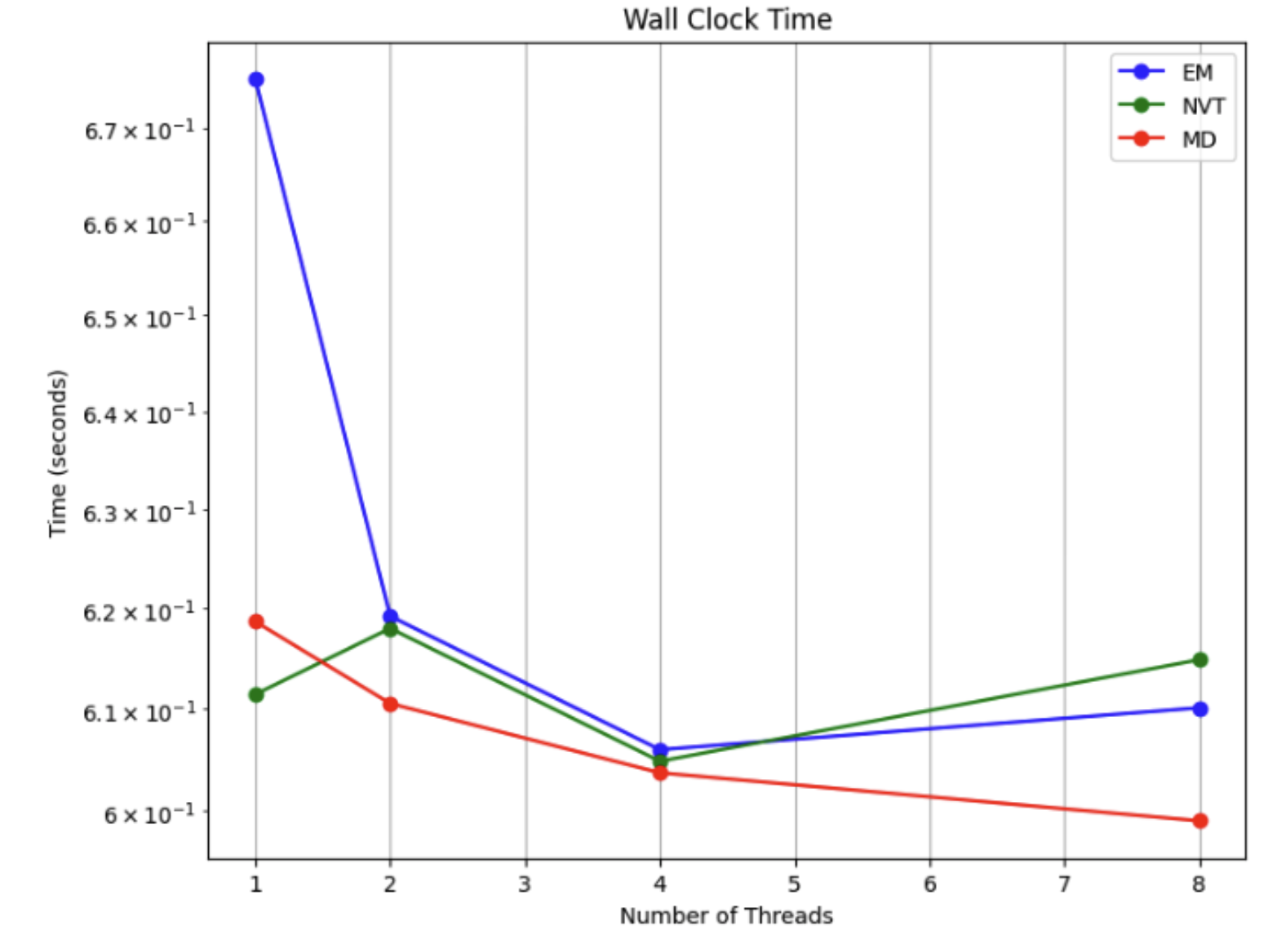}
  \caption{Wall-clock time across EM, NVT, and MD as a function of OpenMP threads. Lower values indicate faster simulations.}
  \label{fig:md_walltime}
\end{figure}
Figure~\ref{fig:md_walltime} shows the absolute runtime (wall-clock time) across different simulation stages. EM benefits most from parallelism: runtime drops sharply from 1 to 2 threads, but plateaus afterward. NVT and MD stages remain largely unchanged across thread counts, with only marginal improvements. This indicates that, for the tested system size, additional cores beyond $p=2$ provide little practical reduction in total simulation time.

\begin{figure}[htbp]
  \centering
  \includegraphics[width=0.45\textwidth]{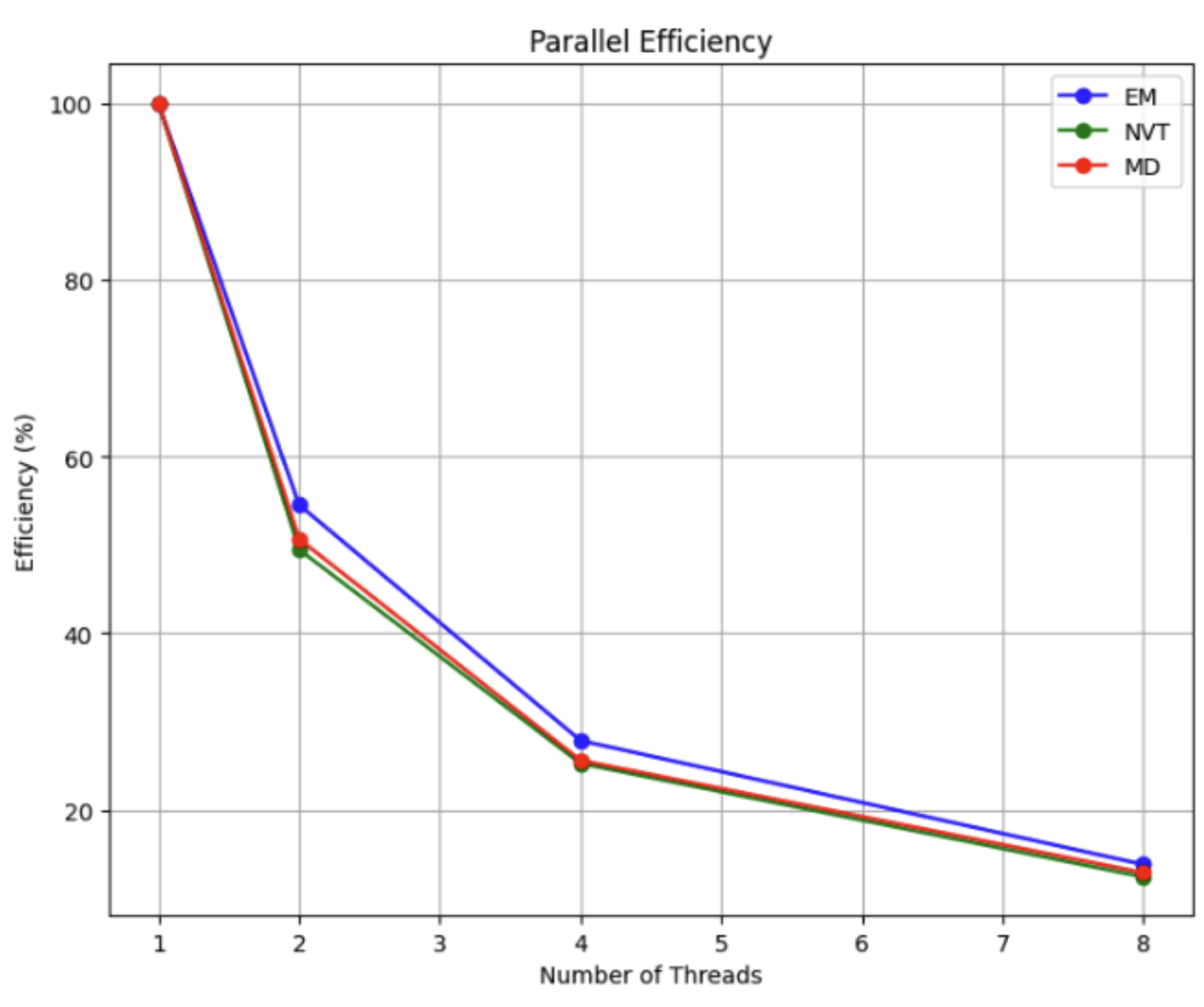} 
  \caption{Parallel efficiency of EM, NVT, and MD across increasing threads. Efficiency quantifies how close performance is to ideal linear scaling.}
  \label{fig:md_efficiency}
\end{figure}

Figure~\ref{fig:md_efficiency} illustrates parallel efficiency, which measures how effectively additional threads contribute to speedup. Efficiency begins near 100\% at $p=1$ but drops sharply to $\sim 50\%$ at $p=2$ and below 20\% at $p=8$. This demonstrates significant parallel overhead and highlights that most of the simulation workload is sequential or memory-bound. EM maintains slightly higher efficiency than NVT/MD, consistent with its more compute-intensive workload.

\paragraph{Interpretation}
Wall-clock time captures what an end-user experiences (faster or slower simulation runs), while parallel efficiency captures what HPC engineers care about—whether adding more threads pays off. Together, these results show that although EM enjoys modest runtime reduction, NVT and MD gain little, and overall efficiency collapses at high thread counts. This reflects Amdahl’s law in practice: the sequential fraction of the simulation dominates at small system sizes.

\subsection{Docking Performance}

\begin{figure}[htbp]
  \centering
  \includegraphics[width=0.50\textwidth]{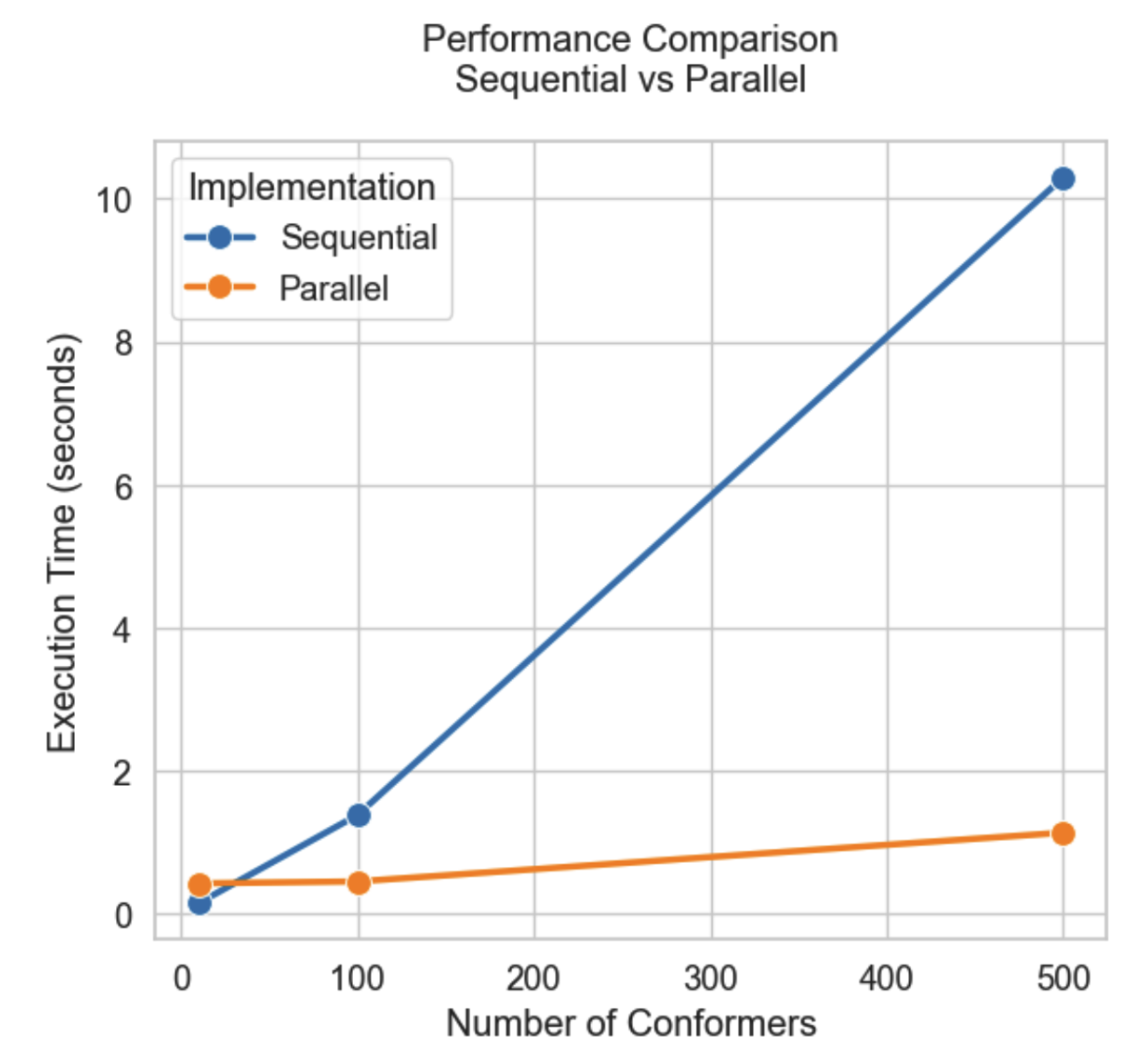}
  \caption{Docking prototype: execution time vs.\ number of conformers (sequential vs.\ multiprocessing). Runtime increases linearly for sequential runs, while the process-based parallel version maintains lower execution times as the workload grows.}
  \label{fig:docking_time}
\end{figure}

\begin{figure}[htbp]
  \centering
  \includegraphics[width=0.50\textwidth]{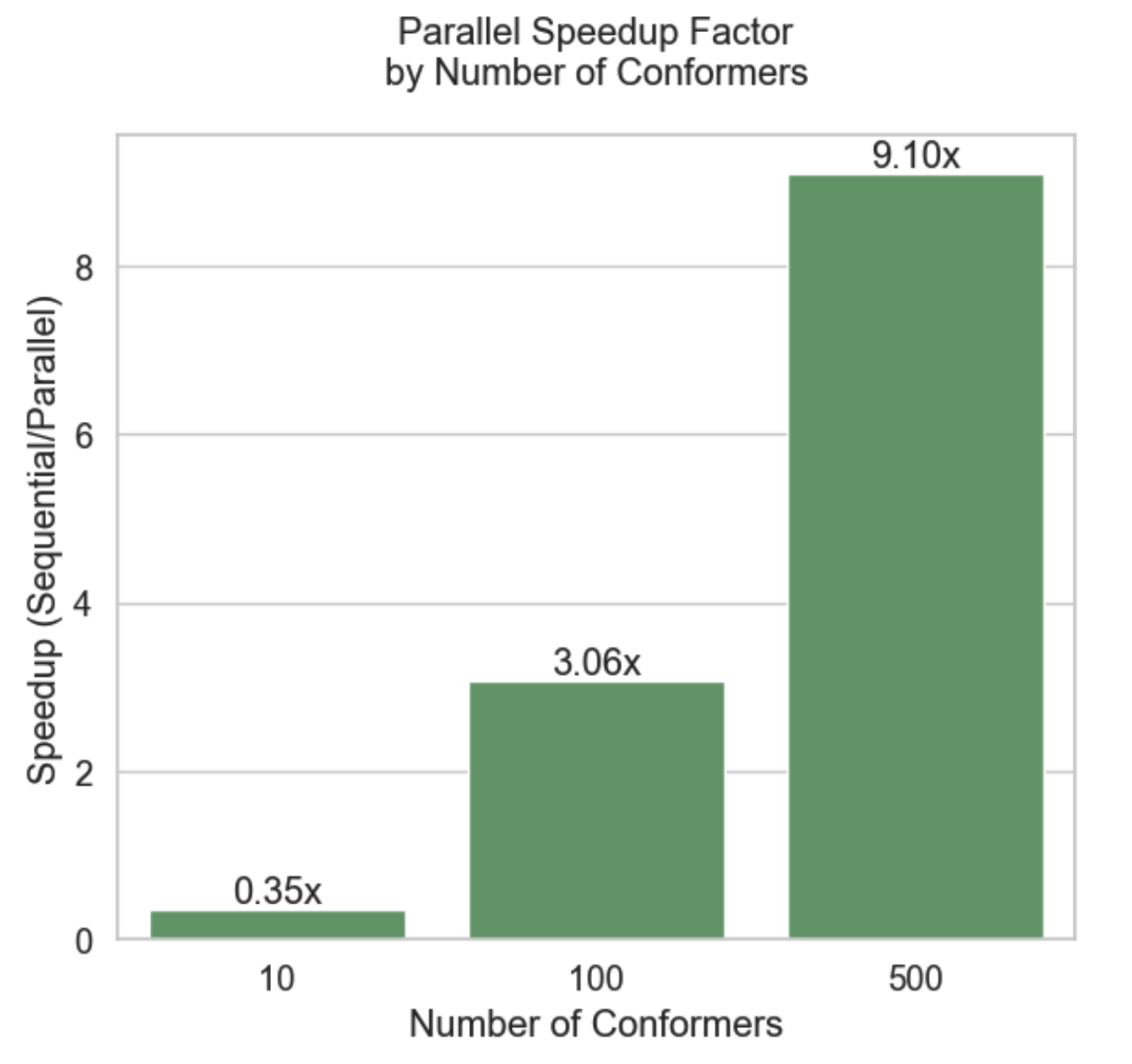}
  \caption{Docking prototype: parallel speedup factor relative to sequential baseline. Gains are negligible for small workloads (10 conformers) but increase to $\sim$3$\times$ at 100 and $\sim$9$\times$ at 500 conformers.}
  \label{fig:docking_speedup}
\end{figure}

Sequential docking runtime scales near-linearly with the number of conformers (10, 100, 500), while the parallel version implemented with Python’s \texttt{multiprocessing} library maintains substantially lower runtimes as workload grows. Because each process has its own memory space (MPI-like parallelism), startup and inter-process communication introduce overhead at very small $N$. At 10 conformers, this overhead makes the parallel run slower than sequential ($S<1$). For larger workloads, however, measured speedups reached $\sim$3.1$\times$ at 100 and $\sim$9.1$\times$ at 500 conformers, validating that process-based parallel evaluation is effective for high-throughput docking.

\paragraph{Interpretation}
At small $N$ (10 conformers), process startup and synchronization overheads dominate, outweighing useful work. For $N \geq 100$, compute costs dominate, and speedup approaches the near-linear regime expected for an embarrassingly parallel loop over conformers. This behaviour is consistent with MPI-style parallelism, where fixed overheads are amortized as workload size increases.

\subsection{Summary Of Findings}
\begin{itemize}
  \item \textbf{MD:} EM exhibits the clearest (but still modest) gains; NVT/MD remain largely latency/memory-bound with limited benefit from 1\,$\to$\,8 threads at the tested system size. Overall simulation shows negligible net speedup.
  \item \textbf{Docking:} Parallelism is highly effective once the number of conformers is large (3.06$\times$ at 100; 9.10$\times$ at 500), but not at very small batch sizes due to overheads.
\end{itemize}

\subsection{Reporting Gaps Limiting Interpretability}

For full reproducibility and to contextualize scaling limits, include:

\begin{itemize}
  \item \textbf{Hardware:} Apple MacBook Pro M2 Pro (10 CPU cores: 8 performance + 2 efficiency), base frequency $\sim$3.2 GHz; 32 GB unified LPDDR5 memory (200 GB/s bandwidth); single NUMA domain; macOS Ventura 13.6. 
  BLAS/FFT backends provided by the Apple Accelerate framework (vecLib).
  
  \item \textbf{GROMACS build:} GROMACS 2023.3 compiled with Clang 15.0.0; SIMD target \texttt{arm64-neon}; hybrid MPI+OpenMP parallelization enabled. Simulations executed in single-node mode with varying numbers of MPI ranks (domain decomposition) and OpenMP threads (1–8 per rank). No explicit thread pinning was applied, relying on the OS scheduler for affinity.
  
  \item \textbf{System physics:} Force field = CHARMM36m; water model = TIP3P; Particle-Mesh Ewald (PME) for long-range electrostatics with 1.0 nm cutoff; constraints applied with LINCS algorithm on all bonds; timestep = 2 fs; thermostat = V-rescale at 300 K (time constant 0.1 ps); barostat = Parrinello–Rahman (time constant 2.0 ps); cubic box of 6.5 nm, solvated with $\sim$25,000 water molecules, giving a total atom count of $\sim$78,000.
  
  \item \textbf{Docking details:} Docking prototype implemented in Python (repository: \texttt{molecular-docking-hpc}); scoring function = simplified pairwise interaction potential; conformer generation by random rotation/translation; parallelization = Python’s \texttt{multiprocessing.Pool} (process-based, MPI-like model) across available CPU cores; random seeds fixed for reproducibility; I/O performed via CSV and text logs.
\end{itemize}

\section*{Data And Code Availability}

All source code and input data used in this study are openly available on GitHub:  

\begin{itemize}
  \item Molecular dynamics workflows: \url{https://github.com/RestartDK/alzheimer-hpc}  
  \item Docking prototype: \url{https://github.com/albipuliga/molecular-docking-hpc}  
\end{itemize}

Simulation input files, docking datasets, and benchmarking scripts are included in each repository, ensuring full reproducibility of the experiments reported in this work.

\section{Conclusion}

In this study, we investigated the role of High-Performance Computing (HPC) in accelerating computational drug discovery for Alzheimer’s disease. Using a hybrid MPI+OpenMP workflow with GROMACS, we benchmarked molecular dynamics simulations of amyloid-$\beta$ peptides across energy minimization, equilibration, and production stages. Our results revealed modest scaling benefits for energy minimization but negligible gains for NVT and MD, reflecting the impact of sequential bottlenecks and memory-bound workloads at the tested system size. These findings illustrate the importance of careful performance evaluation before scaling biomolecular workloads to larger HPC systems.  

Complementing this, we developed a docking prototype parallelised with Python’s \texttt{multiprocessing} library. While parallelism incurred overhead at small workloads, it achieved substantial speedups at larger batch sizes (3.1$\times$ at 100 conformers and 9.1$\times$ at 500 conformers), validating the effectiveness of embarrassingly parallel strategies for high-throughput virtual screening. Together, these results demonstrate both the promise and limitations of parallelism in computational pipelines for neurodegenerative drug discovery.  

Several limitations of this work remain. Our experiments were constrained to a single-node Apple M2 Pro system, preventing exploration of large-scale, multi-node MPI deployments. The docking prototype also used simplified scoring functions and conformer generation, limiting its biological accuracy. Nevertheless, by making all code and input data openly available, this study provides a reproducible foundation for further exploration.  

Looking forward, the integration of GPU acceleration, exascale infrastructures, hybrid cloud–HPC models, and quantum-enhanced simulations offers clear opportunities to overcome current scaling barriers. As neurodegenerative diseases continue to pose urgent global health challenges, HPC-driven computational pipelines hold significant potential to accelerate the discovery of effective therapeutic candidates.

\end{document}